\documentstyle[prc,aps,psfig,preprint,tighten]{revtex}
\begin{document}
\draft
\title{Three-potential formalism for the atomic three-body problem}
\author{Z. Papp \\
Institute of Nuclear Research of the 
Hungarian Academy of Sciences, \\
P.O. Box 51, H--4001 Debrecen, Hungary}
\date{\today}
\maketitle

\begin{abstract}
Based on a three-potential formalism we propose mathematically
well-behaved Faddeev-type integral equations for the atomic three-body 
problem and descibe their solutions in Coulomb-Sturmian space
representation.
 Although the system contains only long-range Coulomb
interactions these equations allow us to reach 
solution by approximating only some auxiliary short-range type
potentials. We outline the method for bound states 
and demonstrate its power in benchmark
calculations. We can report a fast convergence in angular
 momentum channels.
\end{abstract}

\vspace{0.5cm}

The Faddeev equations are the fundamental equations 
of the three-body problems.  Besides giving a unified
formulation, they are superior to the
Schr\"odinger equation in many respects:
in incorporating the boundary conditions,
in treating the symmetries, in handling the correlations, etc..
Nevertheless, their use in
atomic three-body calculations is rather scarce 
\cite{fonseca,schelling,kvitsinsky,hu}, and these calculations 
showed that to reach a reasonable accuracy
many channels are needed. So, the belief spread in the community
that the Faddeev equations, the fundamental
equations of three-body systems, are not well-suited for
treating atomic three-body problems and other techniques
can do a much better job, at least for bound states.
In Refs.\ \cite{fonseca,schelling,kvitsinsky,hu}
the Faddeev equations were used in such a 
form that,  the solution could be reached only
by some kind of approximation on the long-range Coulomb potential,
 and thus the convergence were
ensured only via the square integrability of the bound-state 
wave function.

The aim of this paper is to solve the 
atomic three-body problems by approximating only short-range type
interactions. We invoke a newly established
``three-potential'' formalism and derive such kind  of 
Faddeev-type integral equations which contain only short-range
type interactions as source terms.
We solve the equations by approximating only the short-range type
source terms.
We point out that although we are working with
finite matrices the wave functions possesses correct three-body
Coulomb asymptotics. Finally, as compulsory benchmark cases,
we calculate the helium atom, the positronium ion and the
muonic hydrogen molecule ion, and will observe a fast convergence
with respect to angular momentum channels. 

The ``three-potential'' formalism was designed for
solving nuclear three-body problems in the presence of Coulomb
interaction. The method was presented first in bound-sate
calculations \cite{pzwp} and was extended to below-breakup scattering
calculations \cite{pzsc} where the notion of the ``three-potential''
formalism was also introduced.
In this solution to the nuclear three-body problem, 
in the spirit of the two-potential
formalism, all long-range interactions, i.e.\ all Coulomb 
interactions, are put, 
 \`{a} la Noble \cite{noble}, into the ``free'' Green's operator. 
Invoking again the two-potential formalism,
 the incalculable Faddeev-Noble
Green's operator was linked to a simpler
channel distorted Green's operator. The ``three-potential'' formalism, 
thus consists of repeated applications of the two-potential
formalism, and results in integral equations which
contain only short-range interactions as source terms. 

The Hamiltonian of an atomic three-body system reads 
\begin{equation}
H=H^0 + v_\alpha^C+ v_\beta^C + v_\gamma^C ,
\label{H}
\end{equation}
where $H^0$ is the three-body kinetic energy 
operator and $v_\alpha^C$ denotes the
Coulomb interaction in subsystem $\alpha$. 
We introduce here the usual
configuration-space Jacobi coordinates
$\vec{\xi}_\alpha$ and $\vec{\eta}_\alpha$; 
$\vec{\xi}_\alpha$ is the vector 
connecting the pair $(\beta,\gamma)$ and $\vec{\eta}_\alpha$ is the
vector connecting the center of mass
of the pair $(\beta,\gamma)$ and the particle $\alpha$.
Thus  $v_\alpha^C$ only depends on $\xi_\alpha$,
i.e.\   $v_\alpha^C=v_\alpha^C (\xi_\alpha)$.

The peculiarity of Hamiltonian (\ref{H}) is that all potentials
are of long-range type, and thus the procedure of Refs. \cite{pzwp,pzsc}
cannot be applied without any modification. However,
the physical role of the Coulomb potential is twofold. 
Its long-distance part modifies the asymptotic
motion, while its short-range part strongly correlates the
two-body subsystems.
So, we may split
the Coulomb potential as a sum of long-range and short-range terms,
\begin{equation}
v^C =v^{(l)} +v^ {(s)}.
\end{equation}

Short-range and long-range interactions play entirely
different roles in integral equations. 
While short-range interactions contribute to the source terms, 
long-range interactions should always be put into the 
Green's operator. 
Following Faddeev's procedure \cite{gloeckle}, we 
split the wave function into three components,
\begin{equation}
|\Psi \rangle
=|\psi_{\alpha} \rangle+|\psi_{\beta} \rangle+|\psi_{\gamma} \rangle,
\end{equation}
and, for the components, 
we arrive at the  Faddeev--Noble-type equations
\begin{equation}
|\psi_{\alpha} \rangle= G_\alpha^{(l)} (E ) [ v_\alpha^{(s)}
|\psi_{\beta}\rangle + v_\alpha^{(s)} |
\psi_{\gamma}\rangle ],
\label{feqs}
\end{equation}
with a cyclic permutation for $\alpha, \beta, \gamma$.
The Green's operator is defined as
\begin{eqnarray}
G_\alpha^{(l)} (z) &=& (z-H^0 - v_\alpha^{(l)}
- v_\beta^{(l)} - v_\gamma^{(l)}  - v_\alpha^{(s)})^{-1} \nonumber \\
&=&(z-H^0 - v_\alpha^{C}
- v_\beta^{(l)} - v_\gamma^{(l)} )^{-1}.
\label{g3ldef}
\end{eqnarray}
In the spirit of the ``three-potential'' formalism,
we relate this Green's
operator to the simpler channel distorted 
Green's operator via the resolvent
relation
\begin{equation}
G_\alpha^{(l)} (z) = \widetilde{G}_\alpha (z) + \widetilde{G}_\alpha (z) 
U^\alpha G_\alpha^{(l)} (z), 
\label{lsmulti}
\end{equation}
with $\widetilde{G}_\alpha (z)$ and $U^\alpha$ defined by
\begin{equation}
\widetilde{G}_\alpha (z) = (z-H^0 - v_\alpha^{C}
- u_\alpha^{(l)} )^{-1}
\label{gtildedef}
\end{equation}
and
\begin{equation}
U^\alpha = v_\beta^{(l)} + v_\gamma^{(l)} - u_\alpha^{(l)},
\label{udef}
\end{equation}
respectively.
Here we have introduced the auxiliary potential $u_\alpha^{(l)}$
which is acting in coordinate $\eta_\alpha$ and is required to have
the asymptotic form
\begin{equation}
u_\alpha^{(l)} (\eta_\alpha)
 \sim e_\alpha (e_\beta+e_\gamma) / {\eta_\alpha}
\ \ \ \mbox{as}\ \ \eta_\alpha \to \infty.
\end{equation}
In fact, it is an effective long-range
interaction between the center of mass of the subsystem 
$\alpha$ (with charge $e_\beta+e_\gamma$) and the third particle
(with charge $e_\alpha$).
The importance of $u_\alpha^{(l)}$ lies in the fact that it
asymptotically compensates the Coulomb-tail of
$v_\beta^{(l)} + v_\gamma^{(l)}$ in Eq.\ (\ref{udef}), and thus
 $U^\alpha$  decays faster than the Coulomb interaction
in the two-body channel $\alpha$.
It does not influence the character of the
asymptotic motion, so, although it may contain terms which behave like
$1/\eta_\alpha^2$ as $\eta_\alpha \to \infty$, it can be treated as
short-range interaction. 
So, in Eq.\ (\ref{feqs})
and in Eq.\ (\ref{lsmulti}), which constitute the basic integral
equations of the atomic three-body problem mediated by the
``three-potential'' formalism,
only short-range type interactions 
are appearing as source terms, thus the equations
are mathematically well behaved.
They, similarly to the nuclear Coulomb three-body
problem \cite{pzwp,pzsc},
can be solved in Coulomb-Sturmian (CS) space representation.

The CS functions, which are the Sturm-Liouville solutions of the
hydrogenic problem \cite{rotenberg}, are defined in 
configuration-space for some angular momentum state $l$
 as
\begin{equation}
\langle r|nl\rangle =\sqrt{ \frac{n!}{(n+2l+1)!}}\:
 (2br)^{l+1} \mbox{e}^{-br}L_n^{2l+1}(2br),  \label{basisr}
\end{equation}
where $n=0,1,2,\ldots$, $L$ represents the Laguerre polynomials
and $b$ is a fixed real parameter. 
With the functions $\langle r|\widetilde{nl}\rangle 
=\langle r|nl\rangle/r$ they form a biorthonormal basis.

Since the three-body Hilbert space is a direct sum of two-body 
Hilbert spaces, the appropriate basis in angular momentum 
representation may be defined as
a the direct product 
\begin{equation}
| n \nu l \lambda \rangle_\alpha = 
[ | n l \rangle_\alpha \otimes | \nu
\lambda \rangle_\alpha ]_{L} , \ \ \ \ (n,\nu=0,1,2,\ldots),  
\label{cs3}
\end{equation}
with the CS states of Eq.~(\ref{basisr}). Here $l$
and $\lambda$ denote the angular momenta of coordinates
$\xi$ and $\eta$, respectively, and they are coupled to the total
angular momentum $L$. Now the completeness relation 
takes the form (with
angular momentum summation implicitly included) 
\begin{equation}
{\bf 1} =\lim\limits_{N\to\infty} \sum_{n,\nu=0}^N |
 \widetilde{n \nu l
\lambda} \rangle_\alpha \ \mbox{}_\alpha\langle 
{n \nu l \lambda} | =
\lim\limits_{N\to\infty} {\bf 1}_{N}^\alpha,
\end{equation}
where 
$\langle \xi_\alpha \eta_\alpha |
\widetilde{ n \nu l \lambda }
\rangle_\alpha= {1}/{\xi_\alpha \eta_\alpha} 
\;\langle \xi_\alpha
\eta_\alpha |{\ n \nu l \lambda }\rangle_\alpha$.
It should be noted that in the three-particle 
Hilbert space we can introduce
three equivalent bases belonging to fragmentation 
$\alpha$, $\beta$
and $\gamma$.

In equations (\ref{feqs}) 
we make the following approximation: 
\begin{equation}
|\psi _\alpha \rangle =G_\alpha^{(l)}(E)
[{\bf 1}_N^\alpha v_\alpha^{(s)} 
{\bf 1}_N^\beta |\psi _\beta \rangle +
{\bf 1}_N^\alpha v_\alpha^{(s)} {\bf 1}_N^\gamma 
|\psi _\gamma \rangle ],
\label{feqsapp}
\end{equation}
\noindent
i.e.\ we approximate the short-range potential 
$v_\alpha^{(s)}$ in the three-body
Hilbert space by a separable form 
\begin{equation}
v_\alpha^{(s)}\approx \sum_{n,\nu ,n^{\prime },
\nu ^{\prime }=0}^N|\widetilde{n\nu l\lambda }\rangle _\alpha \;
\underline{v}_{\alpha \beta }^{(s)}
\;\mbox{}_\beta \langle \widetilde{n^{\prime }
\nu ^{\prime }l^{\prime }\lambda
^{\prime }}|,  \label{sepfe}
\end{equation}
where $\underline{v}_{\alpha \beta}^{(s)}=
(1-\delta _{\alpha \beta })\ 
\mbox{}_\alpha \langle n\nu l\lambda |
v_\alpha^{(s)}|n^{\prime }\nu ^{\prime
}l^{\prime }{\lambda }^{\prime }\rangle_\beta$.  
In Eq.\ (\ref{sepfe}) the ket and bra states are defined for different 
fragmentations depending on the
environments of the potential operators in the equations.

Multiplied by the CS states 
$\mbox{}_\alpha \langle \widetilde{n\nu l\lambda }|
$ from the left, Eqs. (\ref{feqsapp}) 
turn into a linear system of homogeneous equations
for the coefficients of the Faddeev 
components $\underline{\psi}_\alpha=\mbox{}_\alpha \langle 
\widetilde{n\nu l\lambda }
|\psi _\alpha \rangle $: 
\begin{equation}
\{ [\underline{G}^{(l)} (E) ]^{-1}-
\underline{v}^{(s)} \}\underline{\psi }=\underline{0}, 
 \label{fep1}
\end{equation}
with 
\begin{equation}
\underline{G}_{\alpha}^{(l)}= 
\mbox{}_\alpha \langle \widetilde{n\nu l\lambda }|
G_\alpha^{(l)}|\widetilde{
n^{\prime }\nu ^{\prime }{l^{\prime }}
{\lambda ^{\prime }}}\rangle _\alpha.
\label{G}
\end{equation}
Eq.\ (\ref{fep1}) is solvable if and only if 
\begin{equation}
\det \{ [\underline{G}^{(l)} (E) ]^{-1}-
\underline{v}^{(s)} \} ={0}. 
\end{equation}
Notice that the matrix elements of the Green's 
operator, which contains all long-range terms,
are needed only between the same partition $\alpha$ whereas 
the matrix elements of the
short range potentials occur only between different 
partitions $\alpha $ and $\beta $.
The latter can be evaluated numerically 
by making use of the transformation 
of Jacobi coordinates \cite{bb}.

The matrix elements (\ref{G}) can be obtained by solving
the Eq.\ (\ref{lsmulti}), which is, in fact, a two-body
multichannel Lippmann-Schwinger equation.
If we perform again the separable 
approximation (\ref{sepfe}) on potential $U^\alpha$,
with the help of the formal solution 
of Eq.\ (\ref{lsmulti}) we may now
express the inverse matrix 
$(\underline{G}^{(l)}_\alpha (E))^{-1}$ as 
\begin{equation}
(\underline{G}^{(l)}_\alpha )^{-1}= 
(\underline{\widetilde{G}}_\alpha )^{-1} -
\underline{U}^\alpha,
\label{glm1}
\end{equation}
where 
\begin{equation}
\underline{\widetilde{G}}_{\alpha} =
 \mbox{}_\alpha\langle \widetilde{n
\nu l \lambda} | \widetilde{G}_\alpha | \widetilde{
 n^{\prime}\nu^{\prime}l^{\prime}{
\lambda}^{\prime}}\rangle_\alpha  \label{gtilde}
\end{equation}
and $\underline{U}^\alpha =
 \mbox{}_\alpha\langle n
\nu l \lambda | U^\alpha | n^{\prime}\nu^{\prime}
l^{\prime}{\lambda}
^{\prime}\rangle_\alpha$.

\begin{figure}[tbp]
\psfig{figure=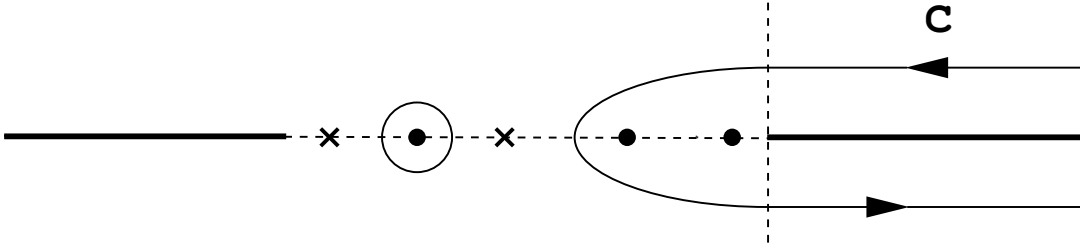,width=15.cm}
\caption{ Contour $C$ for the integral for 
$\widetilde{G}_\alpha (E)$ in case of the three-body bound-state problem
with attractive Coulomb interaction. 
The contour $C$
encircles the continuous and discrete  spectrum of 
$h^C_{\xi_\alpha}$ and avoids the continuous and discrete 
spectrum of $h^{(l)}_{\eta_\alpha}$. The discrete spectrum of 
$h^C_{\xi_\alpha}$ and  $h^{(l)}_{\eta_\alpha}$ is denoted by
dots and crosses, respectively.}
\label{fig2}
\end{figure}

While the latter matrix elements may again be evaluated numerically,
for the calculation of the matrix 
elements in Eq.\ (\ref{gtilde}) we proceed as follows. 
Since we can write the
three-particle free Hamiltonian as a sum of two-particle
free Hamiltonians, 
$H^0=h_{\xi _\alpha }^0+h_{\eta _\alpha }^0$,
the channel distorted Green's operator 
$\widetilde{G}_\alpha$ of Eq.~(\ref{gtildedef})
appears as a resolvent of the sum of two commuting Hamiltonians
$h^C_{\xi _\alpha }=h_{\xi _\alpha }^0+v_\alpha^C$ and
$h^{(l)}_{\eta _\alpha }=h_{\eta _\alpha }^0+u_\alpha^{(l)}$,
which act in different Hilbert spaces. 
Thus, according to
the convolution theorem \cite{bianchi} we can express
the three-body Green's operator $\widetilde{G}_\alpha$
by an integral of two-body Green's operators
\begin{equation}
\widetilde{G}_\alpha (z) = (z-h^C_{\xi _\alpha }-
h^{(l)}_{\eta _\alpha })^{-1} 
= \frac 1{2\pi i}\oint_C  
dz^\prime \,(z^\prime-h^C_{\xi _\alpha })^{-1}
(z-z^\prime-h^{(l)}_{\eta _\alpha })^{-1}. 
 \label{contourint}
\end{equation}
Here the contour $C$ should go, in counterclockwise direction,
 around the continuous and discrete spectrum of $h^C_{\xi _\alpha }$ 
in such a way that the resolvent
of  $h^{(l)}_{\eta _\alpha }$ is analytic in the domain encircled
by $C$ [cf.\ Fig.\ \ref{fig2}]. This condition can only
be fulfilled if we define the  auxiliary potential $u_\alpha^{(l)}$
in such a way that the discrete spectrum of 
$h^{(l)}_{\eta _\alpha }$ does not penetrate into $C$.
The matrix elements (\ref{gtilde}) can be cast into the form 
\begin{equation}
{\underline{\widetilde{G}} (E) } 
 =\frac{1}{2 \pi i} \oint_C dz^\prime  \ 
\mbox{}_\alpha\langle
 \widetilde{ n l}|
(z^\prime - h^C_{\xi_\alpha})^{-1} |
\widetilde{ n^{\prime}{l^{\prime}}}
\rangle_\alpha \  
\mbox{}_\alpha\langle 
\widetilde{ \nu \lambda }|
(E -z^\prime - h^{(l)}_{\eta_\alpha})^{-1} | \widetilde{
\nu^{\prime}{\lambda^{\prime}} }
\rangle_\alpha,  \label{contourint2}
\end{equation}
where both matrix elements occurring in 
the integrand are known from
the two-particle case \cite{cpc}.

After solving Eq.~(\ref{fep1}) for the coefficients $\underline{\psi}$
we can construct the Faddeev components. Considering Eqs.~(\ref{feqsapp}),
 (\ref{lsmulti}) and (\ref{contourint}) we get:
\begin{equation}
|\psi _\alpha \rangle 
=  \bigg[ \frac{1}{2 \pi i} \oint_C dz^\prime  \ 
(z^\prime - h^C_{\xi_\alpha})^{-1} |
\widetilde{ n l}
\rangle_\alpha \  
(E -z^\prime - h^{(l)}_{\eta_\alpha})^{-1} | \widetilde{
\nu \lambda }
\rangle_\alpha \bigg]  \underline{C}_\alpha,  \label{wawefunc}
\end{equation}
where
$\underline{C}_\alpha=
(\underline{\widetilde{G}}_\alpha )^{-1} 
(\underline{G}_\alpha^{(l)} )^{-1} 
[ \underline{v}_\alpha^{(s)} 
 \underline{\psi}_\beta  +
\underline{v}_\alpha^{(s)} 
\underline{\psi}_\gamma]$, and the functions in the integrand 
are also known from two-particle case \cite{cpc}.
So, in this formalism, the Faddeev components appears as a linear
combination of a convolution
integral of Coulomb-like functions. Since in this procedure
we have approximated only short-range type interactions 
$|\psi _\alpha \rangle$ should
possess correct three-body Coulomb asymptotics.

To demonstrate the efficiency of the method we present a few 
benchmark calculations.
As cut-off functions we use
the exceptionally smooth error functions and define
the auxiliary potentials as: 
\begin{equation}
v_\alpha^{(l)}(\xi_\alpha) = 
\mbox{erf}(\omega \xi_\alpha) 
v_\alpha^C (\xi_\alpha),
\end{equation} 
\begin{equation}
v_\alpha^{(s)}(\xi_\alpha) = \mbox{erfc}(\omega \xi_\alpha) 
v_\alpha^C (\xi_\alpha)
\end{equation}
and
\begin{equation}
u_\alpha^{(l)}(\eta_\alpha) =\Lambda \mbox{e}^{-\kappa \eta_\alpha^2}+
 e_\alpha (e_\beta+e_\gamma) \,
\mbox{erf}(\omega  \eta_\alpha) /\eta_\alpha,
\label{upot}
\end{equation}
where $\omega$, $\Lambda$ and $\kappa$ are parameters. In defining
$u_\alpha^{(l)}$, in order to prevent the penetration of the bound-states
of $h^{(l)}_{\eta_\alpha}$  into the contour $C$, 
we have added a repulsive gaussian term. 

We examine
the convergence of the three-body energy with 
increasing  $N$, the number of terms in the
expansion of the short-range potentials 
$v^{(s)}_\alpha$ and $U^{\alpha}$. 
In Table \ref{tabconv} we present results for the binding energy for
helium atom (with infinitely massive core), positronium ion 
($e^- e^+ e^-$)
and the muonic hydrogen molecule ion ($pp\mu^-$). 
For comparison we give the corresponding
result of Faddeev calculations of Ref.\ \cite{schelling,schphd}, where
channel-by-channel comparison is possibble, and of Ref.\
\cite{hu}, where this kind of comparison is not possibble,
and also of results of a very accurate variational calculation
\cite{yeremin}.
In all cases we can observe that, similarly to what we have 
experienced with nuclear
potentials without and with Coulomb interaction \cite{pzwp},
convergence up to 6-7 significant digits is comfortably
achieved with terms up to $N\sim 19$ applied for $n$ and $\nu$, and the
results are in good agreements with well established
benchmark results. 
We can see that with an appropriate choice of $\omega$, 
a rapid convergence is reached in the partial wave expansion.
The convergence in angular momentum channels is much
faster then in earlier works \cite{schelling}.
To overcome the poor convergence, the solution
of the Faddeev equations in total angular momentum representation
was proposed \cite{kvitsinsky,hu}, which, however,
results in three-dimensional equations.
In the light of our results, this seems to be 
superfluous, since the poor convergence is a consequence of
the ill-behavior of the applied form of the Faddeev equations.

It should be noted that in the solution of the Coulomb three-body problem 
the splitting of the Coulomb 
interaction into short- and long-range terms were already suggested
before \cite{merkuriev}. However, this splitting were performed in the
three-body configuration space and the resulted Faddeev differential
equations were used only for scattering state calculations. 
In our case the splitting is performed 
in the subsystem (two-body) Hilbert space and can be effective 
also for bound-state problems.

We have proposed a set of Faddeev and Lippmann-Schwinger
equations for atomic three-body systems using
the newly established ``three-potential'' formalism. 
The Coulomb interactions
were split into long-range and short-range terms and the Faddeev
procedure was applied only to the short-range potentials.
The resulting modified Faddeev equations are mathematically well
behaved since the source terms are of short-range type and
all the long-range interactions are kept in the Green's operator.
This Green's operator was calculated from the channel
distorted Green's operator via two-body multichannel 
Lippmann-Schwinger equations whose kernels are also of short-range
type. Because the channel distorted Green's operator appears as
a resolvent of the sum of two commuting two-body Hamiltonians,
it can be represented as a convolution integral of
the corresponding two-body Green's operators. 
The use of Coulomb-Sturmian functions is essential
as it allows an analytic representation of the
two-body Coulomb Green's operator on the complex energy plane,
which makes straightforward the calculation 
of the convolution integral and
the incorporation of the infinitely many bound
states in an attractive Coulomb potential. In this way the atomic
three-body problem, which contains only long-range Coulomb
interactions, were solved through approximating only some auxiliary
short-range potentials. From this follows that the 
 wave function possesses correct asymptotics.   The extension
of this formalism to below-breakup scattering calculations
is analogous to the nuclear Coulomb case \cite{pzsc}.
We hope, that these unique advantages of this method 
allow us to extend its scope to above-breakup calculations.

This work has been supported by OTKA under Contracts No. T17298 and
T020409.

{}

\newpage

\begin{table}[tbp]
\caption{Convergence of the binding energy
of the $He$ atom, the positronium ion ($e^- e^+ e^-$)
and the muonic hydrogen molecule ion $p p \mu^-$
 with increasing basis for the separable expansion
taking into account angular momentum states up to
$l=0$, $l=2$, $l=4$, $l=6$ and $l=8$.
$N$ denotes the maximum
number of basis states employed for $n$ and $\nu$. 
The values for $He$ atom and  positronium are given in atomic units 
($\hbar=m_e=e^2=1$) and
the values for  $p p \mu^-$ molecule ion are given in modified atomic
units ($\hbar=m_\mu=e^2=1$).
}
\label{tabconv}
\begin{tabular}{rccccc}
& \multicolumn{5}{c}{Angular momentum channels}  \\ 
$N$ & $l=0$ & $l=2$ & $l=4$ & $l=6$ & $l=8$ \\ \hline  
& \multicolumn{5}{c}{$He$ atom: $b=5$, $\omega=1.5$,
 $\Lambda=5$ and $\kappa=2$  }  \\ \hline    
12 & 2.892973  & 2.903778 &  2.903733 & 2.903724 & 2.903723 \\ 
13 & 2.892976  & 2.903780 &  2.903735 & 2.903726 & 2.903725 \\ 
14 & 2.892976  & 2.903781 &  2.903736 & 2.903727 & 2.903725 \\ 
15 & 2.892976  & 2.903781 &  2.903736 & 2.903727 & 2.903726 \\ 
16 & 2.892977  & 2.903781 &  2.903736 & 2.903727 & 2.903725 \\ 
17 & 2.892977  & 2.903781 &  2.903736 & 2.903727 & 2.903725 \\ 
18 & 2.892977  & 2.903781 &  2.903736 & 2.903727 & 2.903725 \\ 
19 & 2.892977  & 2.903781 &  2.903736 & 2.903727 & 2.903725 \\  \hline
 \multicolumn{2}{c}{Faddeev in Ref.\ \cite{schphd}} 
   & 2.9076746287 &  2.9044691278 &  -  &   2.9037959292 \\
\multicolumn{2}{c}{ {with L up to 20 :}} &  & & &
  2.9037224647   \\ \hline
\multicolumn{2}{c}{ Variational in Ref.\ \cite{yeremin}} &  & & &
  2.903724376984   \\ \hline 
& \multicolumn{5}{c}{$e^- e^+ e^-$ ion: $b=0.8$, $\omega=0.2$,
 $\Lambda=5$ and $\kappa=0.3$  }  \\ \hline    
12 & 0.250602 & 0.261753 & 0.261985  & 0.261997 & 0.261998 \\ 
13 & 0.250629 & 0.261761 & 0.261992  & 0.262005 & 0.262006 \\ 
14 & 0.250621 & 0.261763 & 0.261994  & 0.262006 & 0.262008 \\ 
15 & 0.250572 & 0.261758 & 0.261990  & 0.262002 & 0.262003 \\ 
16 & 0.250595 & 0.261762 & 0.261993  & 0.262005 & 0.262006 \\ 
17 & 0.250562 & 0.261760 & 0.261991  & 0.262003 & 0.262004 \\ 
18 & 0.250569 & 0.261760 & 0.261991  & 0.262004 & 0.262005 \\ 
19 & 0.250559 & 0.261760 & 0.261991  & 0.262003 & 0.262005 \\ \hline
\multicolumn{2}{c}{Faddeev in Ref.\ \cite{hu} } &  & & &
  0.2620217   \\ \hline
\multicolumn{2}{c}{ Variational in Ref.\ \cite{yeremin}} &  & & &
  0.2620050702314   \\ \hline 
& \multicolumn{5}{c}{$p p \mu^-$ ion: $b=2.5$, $\omega=0.8$,
 $\Lambda=20$ and $\kappa=0.8$  }  \\ \hline    
12 &   0.462815   & 0.486698 & 0.492922  & 0.494063 & 0.494309 \\ 
13 &   0.458168   & 0.486700 & 0.492921  & 0.494063 & 0.494309 \\ 
14 &   0.455523   & 0.486700 & 0.492922  & 0.494063 & 0.494309 \\ 
15 &   0.453833   & 0.486695 & 0.492918  & 0.494060 & 0.494306 \\ 
16 &   0.452783   & 0.486695 & 0.492918  & 0.494060 & 0.494306 \\ 
17 &   0.452065   & 0.486695 & 0.492918  & 0.494060 & 0.494305 \\ 
18 &   0.451559   & 0.486695 & 0.492918  & 0.494060 & 0.494305 \\ 
19 &   0.451202   & 0.486695 & 0.492918  & 0.494060 & 0.494305 \\ \hline
\multicolumn{2}{c}{Faddeev in Ref.\ \cite{hu} } &  & & &
  0.4943867   \\ \hline
\multicolumn{2}{c}{ Variational in Ref.\ \cite{yeremin}} &  & & &
  0.4943867   \\ 
\end{tabular}
\end{table}

\end{document}